\documentclass[letter]{aa} 
\usepackage{graphicx}

\newcommand{\beq}{\begin{equation}}
\newcommand{\eeq}{\end{equation}}
\newcommand{\beqn}{\begin{eqnarray}}
\newcommand{\eeqn}{\end{eqnarray}}
\newcommand{\lppr}{\stackrel{<}{\scriptstyle \sim}}
\newcommand{\gppr}{\stackrel{>}{\scriptstyle \sim}}

\usepackage{txfonts}
\begin{document}
   \title{Cen A as TeV $\gamma$-ray and possible UHE cosmic-ray source}

   \author{F.M. Rieger
          \inst{1,2}
          \and
           F.A. Aharonian\inst{1,3}}

   \offprints{F.M. Rieger}

   \institute{Max-Planck-Institut f\"ur Kernphysik, Saupfercheckweg 1, 69117 Heidelberg, Germany;
              \email{frank.rieger@mpi-hd.mpg.de}
         \and European Associated Laboratory for Gamma-Ray Astronomy, jointly supported by CNRS and MPG
         \and Dublin Institute for Advanced Studies, 31 Fitzwilliam Place, Dublin 2, Ireland}

   \date{Received 2009; accepted 2009}

 
  \abstract
   {The most nearby active galaxy Cen A has attracted considerable attention as a detected TeV gamma-ray and possible 
   ultra-high energy (UHE) cosmic-ray emitter.}  
   {We investigate the efficiency of particle acceleration close to the supermassive black hole (BH) horizon assuming that
   accretion in the innermost part of the disk occurs in an advection-dominated (ADAF) mode.}   
   {We analyze the constraints on the achievable particle energies imposed by radiative losses and corotation for 
   conditions inferred from observations.}   
   {We show that for an underluminous source such as Cen~A, centrifugally accelerated electrons may reach Lorentz factors 
   of up to $\gamma \sim (10^7-10^8)$, allowing inverse Compton (Thomson) upscattering of ADAF sub-mm disk photons 
   into the TeV regime with an associated maximum (isotropic) luminosity of the order of a few times $10^{39}$ erg/s. 
   Upscattering of Comptonized disk photons is expected to lead to a TeV spectrum $L_{\nu} \propto \nu^{-\alpha_c}$ with 
   a spectral index $\alpha_c \simeq (1.5-1.9)$, consistent with H.E.S.S. results. The corresponding minimum 
   variability timescale could be as low as $r_{\rm L}/c \sim 1$ hr for a typical light cylinder radius of $r_{\rm L} \simeq 5\,
   r_{\rm s}$. While efficient electron acceleration appears to be well possible, protons are unlikely to be accelerated into the 
   extreme UHECR regime close to the central black hole. We argue that if Cen~A is indeed an extreme UHECR emitting source, 
   then shear acceleration along the kpc-scale jet could represent one of the most promising mechanisms capable of pushing 
   protons up to energies beyond 50 EeV.}  
   {}


   \maketitle

\section{Introduction}
    Centaurus A (NGC 5128, Cen~A) is the closest (distance $d \sim 3.4$ Mpc; Israel 1998) and one of the best studied active galaxies. 
    Optically, Cen~A is an elliptical galaxy undergoing late stages of a merger event with a small spiral galaxy. Radio observations 
    detected a complex FR I morphology with a subparsec-scale jet and counter-jet, a one-sided kiloparsec-scale jet, two radio lobes and 
    extended diffusive emission. VLBI observations of the subparsec-scale jet indicate that Cen~A is a non-blazar source with a jet inclination
    angle $i \gppr 50^{\circ}$ (Tingay et al.~1998), or perhaps somewhat smaller if large-scale observations are reliable tracers 
    (Hardcastle et al. 2003). The center of its activity is a supermassive black hole (BH) with a mass inferred to be in the range $m_{\rm BH} 
    \simeq (0.5-1.2) \times 10^8 M_{\odot}$ (Marconi et al. 2006; H\"aring-Neumayer et al. 2006), corresponding to a Schwarzschild scale 
    $r_s \simeq (1.5-3.6) \times 10^{13}$ cm.\\
    Cen~A is the only AGN of the non-blazar type detected at MeV (COMPTEL: Steinle et al. 1998) and GeV energies (EGRET: Sreekumar 
    et al. 1999). The nuclear spectral energy distribution (SED) of Cen~A, as inferred from non-simultaneous data, appears to consist
    of two peaks, one reaching its maximum at several times $10^{13}$ Hz and one peaking around $0.1$ MeV (Chiaberge et al. 2001; 
    Meisenheimer et al. 2007). The SED below 1~GeV has been successfully modeled within a simple jet synchrotron self-Compton (SSC) 
    framework,  assuming Cen~A to be a misaligned BL Lac object (Chiaberge et al. 2001). In the early days of gamma-ray astronomy, a 
    tentative detection ($> 4 \sigma$) of Cen~A at very high energies $>0.3$ TeV was reported during a phase of high X-ray activity in 
    1972-1974 (Grindlay~1975). H.E.S.S. results (Aharonian et al. 2009a) have established Cen~A as a TeV emitting source (the 
    second radio galaxy after M87) with integral VHE flux of $\sim 1\%$ of the Crab Nebula.\\
    Chandra and XMM-Newton observations between 2 and 7 keV indicate that the nuclear X-ray continuum spectrum in Cen~A may  
    consist of both a disk and a jet component, the disk contribution being consistent with a hybrid disk configuration where a standard disk 
    is truncated at $r_t$ and replaced by an ADAF in the inner regions close to the central black hole (Evans et al. 2004; cf. also Pellegrini et 
    al. 2005 and Meisenheimer et al.~2007). A transition to an inner ADAF disk could also explain the lack of a big blue bump UV feature 
    expected in the standard disk scenario (Marconi et al. 2001). The accretion rate for Cen~A seems uncertain by about an order of 
    magnitude, with model-dependent estimates ranging from several $10^{-4}$ to some $10^{-3}\,\dot{m}_{\rm Edd}$. Taking the measured 
    nuclear X-ray luminosity of $L_x \sim 5 \times 10^{41}$ erg/s (Evans et al. 2004) as a robust upper limit to a possible ADAF disk 
    contribution gives $\dot{m} \lppr 0.004 \dot{m}_{\rm Edd}$, which for the canonical viscosity parameter $\alpha \simeq 0.3$ is still below 
    the critical mass accretion rate $\dot{m}_{\rm crit} \simeq 0.3\,\alpha^2\,\dot{m}_{\rm Edd}$ required for a two-temperature ADAF to occur 
    (Narayan et al. 1998; Yi 1999).\\
    If the first SED peak below $10^{14}$ Hz is indeed caused by synchrotron emission from the jet (Chiaberge et al.~2001; Meisenheimer et 
    al.~2007), inverse Compton upscattering in a conventional homogeneous SSC framework is unable to account for VHE $\gamma$-rays 
    in the TeV regime. On the other hand, it could be that the situation for Cen~A is similar to that for M~87, where efficient particle acceleration 
    close to the BH horizon seems to be responsible for the production of the observed TeV $\gamma$-rays (Aharonian et al. 2009b; Rieger 
    \& Aharonian 2008a, henceforth: RA08a). Here we analyze possible VHE characteristics associated with this scenario. We show that 
    for parameters inferred from observations, inverse Compton upscattering of ADAF disk photons by centrifugally accelerated electrons 
    could result in variable ($\sim 1$ hr) VHE $\gamma$-rays with a relatively hard TeV spectrum. We show that the mechanism considered 
    is unable to accelerate protons to ultra high energy cosmic-ray (UHECR) energies, although shear acceleration along the jet could 
    potentially provide a means.    

\section{Cen~A as a TeV $\gamma$-ray source} 
\subsection{Particle acceleration close to the BH horizon}
    Magnetic fields are widely considered to play a key role in the formation of collimated relativistic outflows from compact objects. 
    According to, e.g., MHD scenarios, magnetic flux dragged in by the accretion flow and suitably amplified by dynamo actions in 
    the inner accretion disk can build up a rigidly rotating (quasi-dipolar) disk magnetosphere. Depending on the field-line rotation 
    and the current distribution, the bulk of the plasma may then be centrifugally accelerated along open flux surfaces to relativistic 
    speeds ($\Gamma_b \sim 10$) and become collimated outside the light cylinder $r_{\rm L} \simeq (5-10)\, r_s$ (e.g., Camenzind 
    1995; Fendt 1997). In what follows, we explore an approach where efficient particle acceleration is supposed to occur  
    close to the light cylinder (e.g., Machabeli \& Rogava 1994; Gangadhara \& Lesch 1997; Rieger \& Mannheim 2000; Osmanov et 
    al. 2007).  Within the MHD framework this could occur if, e.g., the longitudinal current becomes smaller than a critical value
    (Beskin \& Rafikov 2000). Let us suppose, thus, that a significant fraction of the electromagnetic energy may be transformed into 
    the kinetic energy of particles by means of magneto-centrifugal effects close to $r_{\rm L}$. Consider for illustration a charged test 
    particle, corotating with the magnetic field and gaining energy while moving outwards (Machabeli \& Rogava 1994). The motion 
    of this particle may be analyzed conveniently in the framework of Hamiltonian dynamics: For length scales much smaller than the 
    curvature radius of the field, the field topology is monopole-like and the associated Hamiltonian $H$ for a particle of rest 
    mass $m_0$ moving along a relativistically, rigidly rotating field-line (angular velocity $\Omega=c/r_{\rm L}=$ constant) becomes 
    a constant of motion
    \beq\label{ham}    
    H=\gamma\,m_0\,c^2 (1-r^2/r_{\rm L}^2)= \mathrm{const.}\,, 
    \eeq where $\gamma=(1-r^2/r_{\rm L}^2-\dot{r}^2/c^2)^{-1/2}$ is the particle Lorentz factor. Thus, as a particle approaches 
    $r_{\rm L}$, its Lorentz factor might increase dramatically (provided corotation holds; however, it cannot become infinite,  
    see below).
    Using Eq.~(\ref{ham}) and the definition of $\gamma$, this can be expressed in terms of a characteristic (minimum) 
    acceleration timescale (cf. RA08a)
    \beq\label{tacc}
      t_{\rm acc}=\frac{\gamma}{\dot{\gamma}} \simeq \frac{1}{2\,\Omega\,
                   \tilde{m}^{1/4}\gamma^{1/2}}\,,
    \eeq valid for $\gamma \gg 1$, where $\tilde{m}=1/(\gamma_0^2\,[1-r_0^2/r_{\rm L}^2]^2)$ depends on the initial injection 
    conditions.

\subsection{Constraints on achievable Lorentz factors}
   Under realistic astrophysical conditions, the acceleration efficiency is constrained by radiation reaction (e.g., inverse Compton in the 
   ambient photon field), the breakdown of the bead-on-the-wire (BW) approximation for a single particle or the bending of the field line 
   with increasing inertia. For a mass accretion rate in the concordant range $(2-20)\times 10^{-4}\,\dot{m}_{\rm Edd}$ (e.g., Marconi et 
   al. 2001; Evans et al. 2004; Pellegrini 2005; Meisenheimer et al. 2007), the ADAF equipartition magnetic field $B_{\rm eq} \propto 
   m_{\rm BH}^{-1/2} r^{-5.4} \dot{m}^{1/2}$ (Yi~1999) becomes $B_{\rm eq}(3\,r_s) \gppr 500$ G, suggesting (radial) jet magnetic field 
   strengths near the light cylinder of $B(r_{\rm L}) \gppr50$ G. In the case of BL Lac objects, transition radii from tens to several hundred 
   $r_s$ have been inferred from observations (e.g., Cao~2003; Wu \& Cao 2005). Similarly, the apparent lack of a standard disk feature 
   in the observed SED of Cen~A seems to indicate as well that $r_t > 10\,r_s$. By assuming that all of the measured X-ray luminosity in 
   Cen~A of $L_x \sim 5 \times 10^{41}$ erg/s is produced by an inner ADAF, the implied energy densities is $U_{\rm ph}\sim L_x/(4\pi\,
   r_t^2 c) \lppr 1$ erg/cm$^3$ for $r_t \gppr 30~r_s$. A similar constraint might be obtained based on the ADAF synchrotron peak luminosity 
   $L_R \sim \nu_s L_{\nu_s} \propto m_{\rm BH}^{6/5}\dot{m}^{4/5} \nu_s^{7/5}$ (Yi 1999) close to $r_{\rm L}$, where $\nu_s(r_{\rm L}) 
   \simeq 7 \times 10^{11}$ Hz, which give $L_R \sim 3 \times 10^{39}$ erg/s, using the fiducial parameters $\dot{m}=5 \times 10^{-4} 
   \dot{m}_{\rm Edd}$, $m_{\rm BH}=10^8 M_{\odot}$, and $r_{\rm L} \sim 5~r_s$ (giving $B(r_{\rm L}) \sim 100$ G), implying a photon 
   energy density $U_{\rm ph}\sim 0.4$ erg/cm$^3$.\\
   (i) Inverse Compton losses (Thomson regime) for electrons, occurring on timescales of $t_{\rm cool} \simeq 3 \times 10^7/[\gamma\,
   U_{\rm ph}]$ s, then imply (see Eq.~(\ref{tacc})) that achievable electron Lorentz factors are limited to 
    \beq\label{InverseCompton}
       \gamma_{\rm max,e}^{\rm IC} \simeq 3 \times 10^8 \sqrt{\tilde{m}}~\left(\frac{10^{14} \mathrm{cm}}{r_{\rm L}}\right)^2 
                                                         \left(\frac{1~\mathrm{erg/cm}^{3}}{U_{\rm ph}}\right)^2\,.
    \eeq
   (ii) For motion along the field synchrotron losses are negligible, while curvature losses, occurring on characteristic timescales of $t_c 
   \simeq 180~R_c^2 (m_0/m_e)~\gamma^{-3}$, do not impose stronger constraints (compared to Eq.~(\ref{InverseCompton})) for curvature 
   radii $R_c \sim r_{\rm L}$.\\
   (iii) The validity of the BW approximation for protons, on the other hand, requires that their characteristic acceleration timescale is always 
   longer than the inverse of the relativistic gyro-frequency (cf. Rieger \& Aharonian 2008b), which constrains the achievable proton Lorentz 
   factors to be 
   \beq\label{breakdown}
       \gamma_{\rm max,p}^{\rm BB} \simeq 10^6\,\tilde{m}^{-1/6}\,
             \left(\frac{B(r_{\rm L})}{50\,\mathrm{G}}\right)^{2/3} 
             \left(\frac{r_{\rm L}}{10^{14}\mathrm{cm}}\right)^{2/3}\,.
    \eeq

\subsection{Expected TeV characteristics}
    The above considerations suggest that electron acceleration close to the central BH in Cen~A may produce Lorentz factors $\gamma 
    \sim 10^7$. This would allow Thomson inverse Compton upscattering of ambient disk photons to TeV energies of $\epsilon \sim \gamma 
    m_e c^2 \simeq 5~(\gamma/10^7)$ TeV. The resultant TeV spectral index at the high energy end then traces, over a decade or so,
    the underlying disk seed photon spectrum (Rieger \& Aharonian 2008b). In the ADAF case, comptonization of thermal synchrotron soft
    photons is expected to lead to a power-law-like tail in the disk spectrum above $\nu_s$, i.e., $L_{\nu} \simeq L_{\nu_s}(\nu/\nu_s)^{-\alpha_c}$, 
    where $\alpha_c = -\ln{\tau_{\rm esc}}/\ln{A}$ depends on the accretion rate (Mahadevan~1997; Yi~1999). For the inferred mass accretion 
    rate of Cen~A, $\dot{m} \simeq (2-20) \times 10^{-4} \dot{m}_{\rm Edd}$, the spectral index becomes $\alpha_c 
    \simeq (1.2-1.9)$ assuming a standard ADAF viscosity coefficient $\alpha =0.3$. This suggests that the upscattered TeV spectrum in 
    Cen~A might in fact be softer than that observed in M87 (see Aharonian et al. 2006). A spectral fit to the VHE ($> 300$ GeV) data 
    obtained by H.E.S.S. indeed suggests a power-law spectral index in the range $1.7 \pm 0.5$ (Aharonian et al. 2009a). A soft spectrum 
    with, e.g.,  $\alpha_c = 1.8$, may then indicate that accretion occurs at the lower end of the concordant interval noted above. 
    Since particle acceleration occurs close to the light surface, the expected minimum TeV variability timescale is $\sim r_{\rm L}/c  \sim 1$ 
    hour.\\   
    Is it possible that TeV photons, produced by inverse Compton upscattering, can escape from the vicinity of the central black hole in 
    Cen~A? In principle, photons of energy E [TeV] will interact most efficiently with target photons in the infrared regime, i.e., of energy 
    $\epsilon_{\rm IR} \simeq$ (1 TeV/E) eV. For the radio luminosity $L_R$ estimated above, the implied infrared ($1$ eV) ADAF disk 
    luminosity is on the order of $L_{\rm IR} \sim 3 \times 10^{39}~(300)^{1-\alpha_c}$ erg/s. The optical depth for 
    $\gamma\gamma$-absorption $\tau_{\gamma\gamma} \simeq (L_{\rm IR}\sigma_{\gamma\gamma})/(4 \pi R_{\rm IR} \epsilon_{\rm IR} c)$ 
    thus becomes
    \beq
     \tau_{\gamma\gamma}(E,R_{\rm IR}) \sim \frac{4}{300^{\alpha_c-1}} \left(\frac{L_{\rm IR}}{3\times 10^{39}~\mathrm{erg/s}}\right) 
                                                                        \left(\frac{r_{\rm L}}{R_{\rm IR}}\right) \left(\frac{E}{1~\mathrm{TeV}}\right)\,,
    \eeq indicating that TeV photons may well escape from the vicinity of the black hole provided the spectrum is sufficiently soft (i.e., for 
    $\alpha \gppr 1.4$). The nuclear (pc-scale) region of Cen~A, on the other hand, is known to be relatively mid-infrared-bright (with 
    luminosity of $\sim 7\times 10^{41}$ erg/s). Yet, if this flux is indeed dominated by emission from a dusty torus of size $\gppr 0.1$ pc 
    (see Radomski et al. 2008), VHE gamma-rays would also be able to escape unabsorbed from the nuclear region of Cen~A.\\
   One can estimate the maximum possible TeV luminosity output $L_{\rm IC} \sim \rho~n~P_c~\Delta V$ based on the Alfv\'en corotation 
   condition $B^2/8\pi \geq n \gamma m_e c^2$, which provides an upper limit for the accelerated particle number density close to 
   $r_{\rm L}$. By noting that for the relevant volume one has $\Delta V =4\pi (\gamma_0/\gamma) r_{\rm L}^3$, that $P_c$ can be 
   approximated by the single particle Thomson power, and that particles are accumulated for $\rho=t_{\rm cool}/t_{\rm acc} 
   \geq 1$ (correcting a typo in RA08a, eq.~12), the maximum possible inverse Compton TeV luminosity is of order
  \beqn
    L_{\rm IC}^{\rm TeV} \sim 10^{39} \left(\frac{B(r_{\rm L})}{50\mathrm{G}}\right)^2 \left(\frac{r_{\rm L}}{1.5 \times 10^{14} 
                                \mathrm{cm}}\right)^2
                                \left(\frac{\gamma_0}{10}\right)^{1/2} \left(\frac{10^7}{\gamma}\right)^{1/2} \mathrm{erg/s}\,, \nonumber 
                                \hspace*{-0.5cm} \\
  \eeqn which is consistent with the H.E.S.S. results of Aharonian et al. (2009a). Since $L_{\rm IC}$ increases almost linearly with black 
  hole mass, a source needs to be sufficiently massive to become detectable by current ground-based instruments. In fact, under normal 
  conditions (i.e., for moderate injection Lorentz factors $\gamma_0 \sim 10$), the inferred TeV output for Cen~A is close to the sensitivity 
  limit ($1\%$ of the Crab flux, with 5$\sigma$ in 50h) of the H.E.S.S. array. Variability studies on hour timescale will thus only be 
  possible with the next-generation CTA telescope project.
 
 \section{On Cen~A as a possible UHECR proton source}
   According to the Pierre Auger results, 4 out of 27 events detected above 57 EeV may be associated with the location of Cen A 
   (Abraham et al. 2007, 2008). If some of these cosmic ray events were indeed associated with Cen A, it seems unlikely that  their 
   energization could occur close to the central BH. Neither a Blandford-Znajek-type nor a centrifugal disk wind scenario seems 
   to allow for that: 
   For example, if the black hole in Cen A rotates and is embedded in a magnetic field of equipartition strength $B_0 \lppr 10^4
   ~(\dot{m}/10^{-2}  \dot{m}_{\rm Edd})^{1/2} (10^8 M_{\odot}/m_{\rm BH})^{1/2}$ G, it could act as a unipolar inductor (membrane 
   paradigm) and induce a voltage difference of the order of
   \beq
   \Phi \sim 4 \times 10^{19} Z a~\left(\frac{m_{\rm BH}}{10^8 M_{\odot}}\right)~\left(\frac{B_0}{10^4 \mathrm{G}}\right)\; \mathrm{V}. 
   \eeq
   However in reality, it seems unlikely that protons might be able to fully tap this potential. Even if the ordered magnetic 
   field component could be as high as $B_0$, screening of the electric field in the plasma-rich AGN environment and curvature 
   losses, for example, are readily expected to reduce achievable proton energies to below $\sim10^{19}$ eV (cf. Levinson 2000; 
   Aharonian et al. 2002). Moreover, as an FR I source, Cen~A is most likely characterized by a Kerr parameter $a$ well below one 
   (e.g., Daly 2009). Centrifugal acceleration of particles in the rotating disk magnetosphere, on the other hand, is limited by the condition 
   for single particle corotation, cf. Eq.~(\ref{breakdown}). While this works well for electrons, efficient centrifugal acceleration of protons 
   to energies far beyond  $10^{17}$ eV is unlikely to occur in the vicinity of the central BH. 
   We finally note, that the required Lorentz factors of $\gamma_p \gppr 5 \times 10^{10}$ (if assumed to be protons) would already imply a 
   gyro-radius $r_{\rm gyro} \simeq 1.5 \times 10^{15} (\gamma_p/5\times 10^{10})~(100~\mathrm{G}/B(r_{\rm L}))$ cm, exceeding the 
   characteristic light cylinder scale for Cen~A.\\
   Furthermore, if radio observations are a reliable tracer of the true fluid speeds in the jet of Cen~A, internal shocks are also probably not 
   able to account for UHECRs beyond 50 EeV. On both VLBA subparsec-scales and VLA scales of hundreds of parsecs, the jet bulk speed 
   seems to be at most mildly relativistic, not exceeding 0.5c by much (Tingay et al. 2001; Hardcastle et al. 2003). If the shocks are 
   internal with characteristic speeds comparable to the relative velocity between layers, shock speeds $\beta_s$ well below $0.5$c are 
   expected. Balancing the minimum timescale for non-relativistic shock acceleration (Bohm limit, e.g. see Rieger et al. 2007) with the 
   timescale for cross-field diffusion then implies, that possible cosmic ray energies are limited to $E_{\rm max} \simeq Z\,e\,B\,r_t\,\beta_s$, 
   where $r_t$ is the (transverse) source size and $B$ the local magnetic field strength. Even if one would allow for some local magnetic field 
   amplification by a factor $\rho$ (e.g., due to shock compression or shear-driven instabilities) and assume that radiative losses could be 
   neglected, possible proton energies are unlikely to exceed
   \beq
     E_{\rm max} \lppr 2\times 10^{19}~ \rho_4~\left(\frac{B_0}{10^4~\mathrm{G}}\right)~\left(\frac{\beta_s}{0.1~\mathrm{c}}\right)\; 
        \mathrm{eV}\,, 
   \eeq where the strength of the local field at $r \sim r_t/\alpha_j$ (where $\alpha_j$ is the jet opening angle) has been expressed as $B 
   \simeq \rho B_0~(r_s/r)$, $\rho_4=\rho/4$, with e.g., $\rho_4=1$ for a strong shock-compressed downstream region. While it seems thus 
   unlikely that protons can be accelerated to $50$ EeV and beyond (at least if no dramatic flaring activity is assumed, cf. Dermer et al. 2009), 
   acceleration to the ankle at $\sim 4 \times 10^{18}$ eV might well be possible.\\
   Hardcastle et al. (2009) recently proposed that efficient stochastic particle acceleration could occur in the giant radio lobes of Cen~A on 
   scales of $R \sim 100$ kpc. In the Bohm limit, the characteristic acceleration timescale for stochastic second order Fermi acceleration is of 
   order $t_{\rm acc} \simeq (c/v_A)^2 (r_{\rm gyro}/c)$ (e.g., Rieger et al. 2007). Comparing this timescale with that for cross-field diffusion, 
   i.e., $t_{\rm df} \sim 3 R^2/(c r_{\rm gyro})$, suggests that proton energies are limited to
   \beq
      E_{\rm max} \simeq 1.6 \times 10^{19} \left(\frac{v_A}{0.1~c}\right)~\left(\frac{R}{100~\mathrm{kpc}}\right)
                               ~\left(\frac{B}{10^{-6}~\mathrm{G}}\right)\;\mathrm{eV}\,.
   \eeq
   Hence, if relativistic turbulent wave speeds can be achieved (of order $v_A \simeq 0.3$c, as assumed in Hardcastle et al. 2009), proton 
   acceleration to energies $5 \times 10^{19}$ eV could become possible. Yet, whether these conditions exist, seems unclear (cf. O'Sullivan 
   et al. 2009). If (part of) the observed X-ray emission from the giant lobes is produced by thermal plasma emission, the implied thermal 
   plasma density would be of order $n_{\rm th} \sim(10^{-5}-10^{-4})$ cm$^{-3}$, i.e., much higher than conducive to fast Alfv\'enic 
   turbulence. For typical equipartition field strengths of $B \sim 10^{-6}$ G, the characteristic Alfv\'en speed $v_A \simeq B/\sqrt{4\pi~n_{\rm 
   th}~m_p} \lppr c/500$ would be much lower than required.\\
   Based on the above considerations, it seems questionable whether protons could indeed be accelerated to energies beyond $10^{19}$ 
   eV in Cen A. However, one possible mechanism that may help particles to reach beyond this limit could be shear acceleration along the 
   large-scale jet (Rieger \& Duffy 2004). Strong internal jet stratification (e.g., a fast spine surrounded by slower moving layers) is naturally 
   expected for jets interacting with their environments or launched by different processes (e.g., disk and black-hole driven wind), and indeed 
   seen in hydrodynamical simulations (e.g., Meliani \& Keppens 2007; Aloy \& Mimica 2008). In the case of Cen~A, there are some 
   observational hints, e.g., limb-brightening in the X-ray jet, the higher energy emission being more confined to the jet axis, longitudinal 
   magnetic field polarization in the large-scale jet (Kataoka et al. 2006; Kraft et al. 2002; Hardcastle et al. 2003) that might be indicative of  
   internal jet stratification. Energetic particles scattered across such a shear flow can sample the kinetic difference in the flow and be accelerated 
   to higher energies. The timescale for particle acceleration in a continuous shear flow is inversely proportional to the particle mean free path 
   $t_{\rm acc} \propto 1/\lambda$ (Rieger \& Duffy 2004). In order to compete with radiative losses, and in the case of Cen~A with advection 
   along the jet, high energy seed particles are thus needed to ensure efficient particle acceleration. 
   Assuming a gyro-dependent particle mean free path $\lambda \sim r_{\rm gyro}$, the minimum acceleration time scale in a linearly decreasing 
   velocity shear profile is on the order of $t_{\rm acc} \simeq 3 (\Delta r)^2 c/[\lambda~\gamma_b^4 (\Delta v_z)^2] \sim 6~(\Delta r)^2/(r_{\rm gyro} c)$ 
   assuming $v_z \sim 0.5$c (see Rieger \& Duffy 2004). Advection along the jet occurs on a characteristic timescale $t_a \sim l_j/v_z \sim 10^{12}$ 
   sec, where $l_j \sim 4.5$ kpc is the (projected) X-ray jet length (Hardcastle et al. 2007). The observed kpc radio jet slightly expands and has 
   a transverse width of $\sim 1$ kpc close to its end (Kraft et al. 2002). Protons injected at the crossover to the kpc-scales, for example, would 
   thus need seed energies of $E \sim (1-5) \times 10^{18}$ eV to ensure that shear acceleration can operate efficiently (using, e.g., $\Delta r \sim  
   0.2$ kpc, $B\sim 10^{-4}$ G). Such energetic seed particles may well be provided by e.g. shock acceleration. In this case, protons escaping from 
   the inner shock region and diffusing into the outer shear layers will naturally experience an additional increase in energy. In 3d-hydrodynamical 
   simulations (Aloy et al. 1999), the shear layer is found to broaden with distance along the jet, eventually becoming comparable to the beam width. 
   If one thus takes $\Delta r \sim 1$ kpc near to the jet end (we note that the full width of the turbulent shear layer may be even larger than that seen 
   in the radio), the classical Hillas confinement condition would constrain maximum protons energies to be $E  \simeq 10^{20} (B/10^{-4}\mathrm{G})$ 
   eV. This value might be somewhat higher, if the magnetic field is amplified in the shear (e.g., due to Kelvin-Helmholtz-driven shear instabilities, see 
   Zhang et al. 2009). On the other hand, for smaller transverse shear widths, non-gradual shear acceleration (Ostrowski 1998; Rieger \& Duffy 2004) 
   is expected to take over earlier, leading to similar maximum energy considerations but different spectral characteristics.       
 
\section{Conclusions}
    If Cen~A hosts a radiatively inefficient inner accretion disk, efficient electron acceleration in its BH-jet magnetosphere and inverse Compton 
    upscattering of  ambient disk photons to the TeV regime appears to be possible. After M87, Cen~A could thus be another primary candidate 
    source in which VHE processes close to the black hole may allow a fundamental diagnosis of its immediate environment, offering important 
    insights into the central engine in AGNs.\\ 
    Whether Cen~A indeed proves to be an UHECR emitting source may require confirmation by additional observations. From a theoretical 
    point of view and based on conventional acceleration concepts, it seems challenging to account for a possible production of UHECR protons 
    beyond $10^{19}$ eV. On the other hand, as shown above, shear acceleration along the large-scale jet in Cen~A may help to overcome 
    this problem by increasing the energy of shock-accelerated seed particles by a factor of some tens. If this would be the case, spectral changes 
    in the cosmic ray spectrum might be partly due to the operation of a new acceleration mechanism and not simply propagation effects.

\begin{acknowledgements}
    Discussions with B. Reville, A. Taylor, C. Fendt and V. Beskin, and useful comments by the referee are gratefully acknowledged.
\end{acknowledgements}


\begin{thebibliography}{}
 
  \bibitem[2007]{abr07}
     Abraham, J., et al. (Pierre Auger Collaboration) 2007, Science 318, 938

  \bibitem[2008]{abr08}
     Abraham, J., et al. (Pierre Auger Collaboration) 2008, APh 29, 188
        
  \bibitem[2002]{aha02}
       Aharonian, F.A. et al. 2002, Phys. Rev. D 66, 023005       
       
  \bibitem[2006]{aha06} 
       Aharonian, F., et al. (H.E.S.S. Collaboration) 2006, Science 314, 1424
       
  \bibitem[2009a]{aha09a} 
       Aharonian, F., et al. (H.E.S.S. Collaboration) 2009a, ApJ 695, L40   
       
  \bibitem[2009b]{aha09b}      
     Aharonian, F., et al. (H.E.S.S. Collaboration) 2009b, Science 325, 444 

  \bibitem[1999]{alo99}
          Aloy, M.A. et al. 1999, ApJ 523, L125

   \bibitem[2008]{alo08}
	Aloy, M.A., Mimica, P. 2008, ApJ 681, 84

  \bibitem[2000]{bes00}
        Beskin, V.S., Rafikov R.R. 2000, MNRAS 313, 433

   \bibitem[2001]{chi01}
       Chiaberge M., Capetti A., Celotti A. 2001, MNRAS 324, L33
       
   \bibitem[1995]{cam95}
       Camenzind, M. 1995, RvMA 8, 201
  
  \bibitem[2003]{cao03}
       Cao, X. 2003, ApJ 599, 147
     
 \bibitem[2009]{dal09}
       Daly, R.A. 2009, ApJ 691, L72
     
 \bibitem[2009]{der09}
      Dermer, C.D. et al. 2009, NJP 11, 065016

  \bibitem[2003]{dim03}
       Di Matteo, T. et al. 2003, ApJ 582, 133

  \bibitem[2004]{eva04}
       Evans, D.A. et al. 2004, ApJ 612, 786

  \bibitem[1997]{fen97}
       Fendt, C. 1997, A\&A 319, 1025

  \bibitem[1997]{gan97}
       Gangadhara, R.T., Lesch, H. 1997, A\&A 323, L45

  \bibitem[1975]{gri75}
       Grindlay, J.E. et al. 1975, ApJ 197, L9

   \bibitem[2006]{hae06}
       H\"aring-Neumayer, A. et al. 2006, ApJ 643, 226

  \bibitem[2003]{har03}
       Hardcastle, M.J. et al. 2003, ApJ 593, 169

  \bibitem[2007]{har07}
       Hardcastle, M.J. et al. 2007, ApJL 670, L81

 \bibitem[2009]{had09}
       Hardcastle, M.J. et al. 2009, MNRAS 393, 1041 
       
  \bibitem[1998]{isr98}
       Israel, W. 1998, A\&A Rev. 8, 237
       
   \bibitem[2006]{kat06}
       Kataoka J., et al. 2006, ApJ 641, 158      
       
  \bibitem[2002]{kra02}       
       Kraft, R.P. et al. 2002, ApJ 569, 54 
       
  \bibitem[2000]{lev00}
       Levinson, A. 2000, Phys. Rev. Lett. 85,912

  \bibitem[1994]{mac94}
       Machabeli, G.Z., Rogava, A.D. 1994, Phys. Rev. A 50, 98 
       
  \bibitem[1997]{mah97}
       Mahadevan, R. 1997, ApJ 477, 585

 \bibitem[2001]{mar01}
       Marconi, A. et al. 2001, ApJ 549, 915

 \bibitem[2006]{mar06}
       Marconi, A. et al. 2006, A\&A 448, 921

  \bibitem[2007]{mei07}
       Meisenheimer, K. et al. 2007, A\&A 471, 453

  \bibitem[2007]{mel07}
       Meliani Z., Keppens, R. 2007, A\&A 475, 785
  
  \bibitem[1998]{nar98}
       Narayan, R., Mahadevan R., Quataert, E. 1998, in:
       Theory of Black Hole Accretion Disks, eds. M.A. Abramowicz et al.,
       Cambridge, p. 148

 \bibitem[2007]{osm07}
       Osmanov, Z., Rogava, A., Bodo, G. 2007, A\&A 470, 395

 \bibitem[1998]{osr98}
       Ostrowski, M. 1998, A\&A 335, 134

 \bibitem[2009]{osu09}
       O'Sullivan, S., Reville, B., Taylor, A. 2009, MNRAS in press (arXiv:0903.1259)

  \bibitem[2005]{pel05}
       Pellegrini, S. 2005, ApJ 624, 155
       
\bibitem[2008]{rad08}
      Radomski J.T., et al. 2008, ApJ 681, 141
       
   \bibitem[2000]{rie00}
       Rieger, F.M., Mannheim, K. 2000, A\&A 353, 473 (RM00)
       
\bibitem[2004]{rie04}
       Rieger, F.M., Duffy, P. 2004, ApJ 617, 155
 
 \bibitem[2007]{rie07}
       Rieger, F.M., Bosch-Ramon, V., Duffy, P. 2007, Ap\&SS 309, 119

  \bibitem[2008]{rie08a}
       Rieger, F.M., Aharonian F.A. 2008a, A\&A 479, L5 (RA08a)

  \bibitem[2008]{rie08b}
     Rieger, F.M., Aharonian F.A. 2008b, IJMPD 17, 1569 
     
  \bibitem[1999]{sre99}
       Sreekumar, P. et al. 1999, APh 11, 221
       
  \bibitem[1998]{ste98}
       Steinle, H. et al. 1998, A\&A 330, 97

  \bibitem[1998]{tin98}
       Tingay, S.J. et al. 1998, AJ 115, 960
       
 \bibitem[2001]{tin01}
       Tingay, S.J. et al. 2001, AJ 122, 1697

 \bibitem[2005]{wu05}
        Wu, Q., Cao, X. 2005, ApJ 621, 130

  \bibitem[1999]{yi99}
        Yi, I. 1999, in: Astrophysical Disks, ASP Conf. Ser. 160,
        eds. J.A. Sellwood, J. Goodman, p.~279

\bibitem[2009]{zha09}
	Zhang, W., MacFadyen, A., Wang, P. 2009, ApJ 692, L40

\end{thebibliography}
\end{document}